\documentclass[a4paper,twocolumn,
english,aps,prb,floatfix,showpacs]{revtex4}
\usepackage[T1]{fontenc}
\usepackage[latin1]{inputenc}
\usepackage{amsmath}
\usepackage{babel}
\usepackage{graphics}
\usepackage{amssymb}

\begin{document}
 
\title{On locations and properties of the multicritical point of Gaussian and 
$\pm J$ Ising spin glasses 
}

\author{S.L.A. \surname{de Queiroz}}

\email{sldq@if.ufrj.br}

\affiliation{Instituto de F\'\i sica, Universidade Federal do
Rio de Janeiro, Caixa Postal 68528, 21941-972
Rio de Janeiro RJ, Brazil}

\date{\today}

\begin{abstract}
We use transfer-matrix and finite-size scaling methods to
investigate the location and 
properties of the multicritical point of 
two-dimensional Ising spin glasses on square, triangular and  honeycomb  
lattices, with both binary and Gaussian disorder distributions. 
For  square and triangular lattices with binary disorder, the estimated
position of the multicritical point is in numerical agreement with
recent conjectures regarding its exact location. For the remaining four
cases, our results indicate disagreement with the respective versions
of the conjecture, though by very small amounts, never exceeding $0.2\%$. 
Our results for: (i) the correlation-length exponent $\nu$ governing
the ferro-paramagnetic transition; (ii) the critical domain-wall energy 
amplitude $\eta$;
(iii) the conformal anomaly $c$; (iv) the finite-size susceptibility exponent
$\gamma/\nu$; and (v) the set of multifractal exponents $\{ \eta_k \}$ associated to
the moments of the probability distribution of spin-spin correlation functions
at the multicritical point, are consistent with universality as regards lattice 
structure and disorder distribution, and in good agreement with existing estimates.
\end{abstract}
\maketitle
\section{INTRODUCTION}
\label{intro}
In this paper we investigate quenched random-bond Ising spin - $1/2$ models on regular
two-dimensional lattices, namely square [SQ], triangular [T], and honeycomb [HC]. 
For suitably low concentrations of antiferromagnetic bonds, 
it is known that such systems exhibit ferromagnetic order at low temperatures.  
We consider only nearest-neighbor couplings  $J_{ij}$, with strengths
extracted from identical, independent probability distribution functions (PDFs).
We specialize to the following two forms for the latter:
\begin{eqnarray}
P(J_{ij})= p\,\delta (J_{ij}-J_0)+ (1-p)\,\delta (J_{ij}+J_0)\ \qquad  (\pm J)\ ;
\nonumber \\
P(J_{ij})= 
\frac{1}{\sqrt{2\pi}\sigma}\,\exp\left(-\frac{(J_{ij}-J_0)^2}{2\sigma^2}\right)\quad 
\ {\rm (Gaussian)}\  .\quad
\label{eq:1}
\end{eqnarray}
Our units are such that $J_0 \equiv 1$ in the former case, and $\sigma \equiv 1$
in the latter. A critical line on 
the $T - p$ ($\pm J$), or $T - J_0$ (Gaussian), plane separates paramagnetic and 
ferromagnetic phases; a spin-glass phase for comparable amounts of plus and minus
couplings is absent here,
on account of the systems under consideration being two-dimensional.

For general space dimensionality $d \geq 2$ there is a second line of 
interest on  the temperature-disorder plane, along which the internal energy
has a simple analytic expression, and several exact results have been derived, 
known as the {\it Nishimori line} (NL)~\cite{nish81,nish01}. The shape of the
NL is known exactly, and given by
\begin{eqnarray}
e^{-2/T} = \frac{1-p}{p}\qquad\ (p> \frac{1}{2})
\qquad  (\pm J)\ ;
\nonumber \\
T = \frac {1}{J_0}\qquad\qquad {\rm (Gaussian)}\ .
\label{eq:2}
\end{eqnarray}
The intersection of the ferro-paramagnetic boundary with the NL is a multicritical 
point\cite{ldh88}, the {\it Nishimori point} (NP). 
A conjecture regarding the possibly exact location of the NP has been put 
forward, which invokes the effects of duality and gauge symmetry arguments
on the replicated partition function of quenched random $Z_q$ 
models~\cite{nn02,mnn03,tn04}. With further extensions
to consider non self-dual lattices~\cite{tsn05,no06}, numerically exact
predictions have been produced, for the $Z_2$ (Ising) model, for all lattices and 
interaction distributions considered here. Versions of the
conjecture adapted for hierarchical lattices have been considered as 
well~\cite{hb05}. 

Locations of the NP predicted by the conjecture 
generally agree very well with results obtained
by other means. However, the remaining discrepancies provide compelling evidence
that, at least in some cases, the conjecture may not be exact. First, on the 
SQ lattice, several very accurate numerical estimates for the $\pm J$ coupling
distribution place the 
conjectured location~\cite{nn02}, $p_c=0.889972\dots$, outside the corresponding 
error bars (though it differs from the central 
value typically by less than $0.1\%$). One has:
$p_c=0.8906(2)$~(Refs.~\onlinecite{hpp01,php06}), 
$0.8907(2)$~(Ref.~\onlinecite{mc02a}), 
$0.89081(7)$~(Ref.~\onlinecite{has08}). For a Gaussian distribution, the conjecture
gives~\cite{nn02,mnn03} $J_{0c}=1.021770$, while Ref.~\onlinecite{php06} finds
$J_{0c}=1.02098(4)$. 
Second, it has been shown that the  exact  renormalization-group
solution for three pairs of mutually dual hierarchical lattices disagrees 
with the pertinent form of the conjecture, by up to $2\%$~\cite{hb05}.

Very recently, these issues have been addressed via the proposal of an improved
conjecture, first applied to hierarchical lattices~\cite{onb08}, and later
extended to regular ones~\cite{ohzeki08}. Broadly, this corresponds to 
considering duality properties applied to a (usually small) cluster of sites 
on the lattice under examination~\cite{onb08,ohzeki08}, as opposed to the 
original conjecture which deals only with the partition function of a
single bond (the principal Boltzmann factor)~\cite{nn02}.
The improved conjecture predicts the location of the NP to be well within
the error bars of recent numerical work for the SQ $\pm J$ case:
an average over four slightly differing implementations gives $p_c=0.89079(6)$, 
though so far disagreement persists for the Gaussian distribution,
as the improved estimate is $J_{0c}=1.021564$~\cite{ohzeki08}.
For hierarchical lattices, the gap between conjecture and exact
renormalization-group solutions has essentially been bridged by the new 
approach~\cite{onb08}.

Existing numerical results for T and HC lattices ($\pm J$ distribution 
only)~\cite{tsn05,dq06} broadly agree with an early form of the original 
conjecture, applicable to pairs of dual lattices~\cite{tsn05}:
with the binary entropy 
\begin{equation}
H(p) \equiv -p\,\log_2 p - (1-p)\,\log_2 (1-p)\ ,
\label{eq:h(p)}
\end{equation}
it is predicted that, for a pair of mutually-dual lattices 1 and 2,
\begin{equation}
H_{12} \equiv H(p_{1c}) + H(p_{2c}) =1\ .
\label{eq:earlyconj}
\end{equation}
Ref.~\onlinecite{tsn05} finds $p_c=0.835(5)$ and $0.930(5)$, respectively for
T and HC, which implies $0.981 < H_{12} < 1.042$; these estimates
were refined in Ref.~\onlinecite{dq06} to $p_c=0.8355(5)$~[T] and $0.9325(5)$~[HC], 
giving $H_{12}= 1.002(3)$. 

Further developments~\cite{no06,ohzeki08} enabled the
production of pairs of individual predictions 
(always obeying Eq.~(\ref{eq:earlyconj}), 
with a suitably-adapted form of Eq.~(\ref{eq:h(p)}) for the Gaussian case).
In  the framework of the original conjecture, these are: $p_c=0.835806$ [T]
and $0.932704$ [HC] ($\pm J$)~\cite{no06}; $J_{0c}=0.798174$ [T] and
$1.270615$ [HC] (Gaussian)~\cite{ohzeki08}. For the improved conjecture 
($\pm J$ only), two slightly differing implementations give the pairs: 
$p_c=0.835956$ [T] and $0.932611$ [HC];  $p_c=0.835985$ [T]
and $0.932593$ [HC]~\cite{ohzeki08}. 

Here, we numerically estimate the location and critical properties of the NP on
the T and HC lattices. For the $\pm J$ case, we refine
the results given in Ref.~\onlinecite{dq06}, checking our data against the more
stringent predictions of Refs.~\onlinecite{no06,ohzeki08}; for the Gaussian
distribution, we are not aware of any existing results, apart from those
given in Ref.~\onlinecite{ohzeki08} for the conjectured location of the NP.
For completeness, and to provide consistency checks of our methods,
we revisit the SQ lattice problem, investigating both distributions.

We apply numerical transfer-matrix (TM) methods to the spin--$1/2$ random-bond 
Ising model, on strips of SQ, T, and HC lattices, of widths $4 \leq N \leq 
14$ sites (SQ), $4 \leq N \leq 13$ sites (T) and $4 \leq N \leq 14$ sites 
(even values only, HC). 
We take long strips, usually of length $M=2 \times 10^6$ columns (pairs of 
columns for  HC, because two iterations of the TM are needed to restore 
periodicity). For each of the quantities evaluated here, averages $\langle {\cal Q} 
\rangle$ are taken 
over, and fluctuations $\langle \Delta {\cal Q} \rangle_{\rm rms}$
calculated among, $N_s$ independently-generated samples, 
each of length $M$. As discussed extensively elsewhere~\cite{dqrbs96},
the sample-to-sample fluctuations $\langle \Delta{\cal Q} \rangle_{\rm rms}$ 
vary with $M^{-1/2}$, and are 
essentially $N_s$--independent, provided that $N_s$ is not very small. 
The averaged values $\langle {\cal Q} \rangle$ themselves still fluctuate slightly upon 
varying $N_s$, but the corresponding fluctuations $\Delta \langle {\cal Q} \rangle$
die down with increasing $N_s$.
We found that, for $M$  as above, making $N_s=10$ already gives $\Delta \langle {\cal 
Q} \rangle / \langle \Delta {\cal Q} \rangle_{\rm rms} \lesssim 0.1$, thus this
constitutes an adequate compromise between accuracy and CPU time expense. 
Typical upper bounds for $\langle \Delta {\cal Q} \rangle_{\rm rms}/\langle {\cal Q} 
\rangle$ are $10^{-4}$ for free energies,  $10^{-3}$ for domain-wall energies
(see Section~\ref{sec:2} below for definitions). 

We scanned suitable intervals of $p$ or $J_0$ along the NL, spanning 
conjectured and (when available) numerically-calculated positions
of the NP, as shown in Table~\ref{tpar}. 
For a given lattice and interaction distribution,
we took samples at $N_p=N_p (N)$ equally-spaced positions for each lattice width $N$, 
generally starting with $N_p \geq 18$ 
for small $N$, and  decreasing to $N_p=9$ for $N \geq 8$, giving the totals
denoted by ${\cal N}_p = \sum_N N_p(N)$ in Table~\ref{tpar}.     
\begin{table}
\caption{\label{tpar}
Intervals $\Delta p$, $\Delta J_0$ scanned along the NL in our calculations,
for lattices and coupling distributions
[binary ($\pm J$) and Gaussian (G)] specified in column 1. ${\cal N}_p$
gives total number of pairs $(p,N)$ or $(J_0,N)$ at which quantities of interest 
were calculated. See text.
}
\vskip 0.2cm
\begin{ruledtabular}
\begin{tabular}{@{}lcc}
Type &  $\Delta p$, $\Delta J_{0}$ & ${\cal N}_p$ \\
SQ, $\pm J$ & $0.8868$ -- $0.8948$ & $123$ \\
SQ, G &  $1.00125$ -- $1.03875$ & $140$\\
T, $\pm J$ & $0.830$ --  $0.842$ & $126$ \\
HC, $\pm J$ & $0.9266$ -- $0.9386$  & $86$ \\
T, G &  $0.7794$-- $0.8169$ & $134$ \\
HC, G &  $1.254$ -- $1.287$ & $94$ \\
\end{tabular}
\end{ruledtabular}
\end{table}

The Mersenne Twister random-number generator~\cite{mt} 
was used in all calculations described below.  
In all calculations pertaining to
the $\pm J$ disorder distribution, a canonical ensemble was used, i.e. for a 
given  nominal concentration $p$ of positive bonds, these were drawn from
a reservoir initially containing $\alpha_i pNM$ units ($\alpha_i=2,3,3$ 
respectively for $i=$ SQ,T, HC). This way, one ensures that
fluctuations in calculated quantities are considerably smaller than if
a grand-canonical implementation were used~\cite{ahw98,php06}.

In Sec.~\ref{sec:2}, domain-wall energies are computed, and their finite-size
scaling allows us to estimate both the location of the NP along the NL,
and the correlation-length index, $y_t \equiv 1/\nu$ which governs the spread
of ferromagnetic correlations upon crossing the ferro-paramagnetic phase boundary.
The conformal anomaly, or central charge, is evaluated in Sec.~\ref{sec:2a}.
In Sec.~\ref{sec:susc}, uniform susceptibilites are calculated, and the associated
exponent ratio $\gamma/\nu$ is evaluated (for Gaussian coupling distributions only). In 
Sec.~\ref{sec:cf}, we specialize to T and HC lattices,  with Gaussian disorder 
distributions, and investigate the moments of assorted orders of the
probability distributions of spin-spin correlation functions. 
Finally, in Sec.~\ref{conc}, concluding remarks are made.

\section{Domain-wall scaling}
\label{sec:2}
For a strip of width $L$, in lattice parameter units, of a two-dimensional
spin system, the domain-wall free energy $\sigma_L$ is the free energy per 
unit length, in units of $T$, of a seam along the full length of the strip.
For Ising spins, $\sigma_L =f^A_L-f^P_L$, with $f^P_L$ ($f^A_L$) 
being the corresponding free energy for a strip with periodic (antiperiodic) 
boundary conditions across. Within a TM description of disordered
systems, $\sigma_L = -\ln (\Lambda_0^A / \Lambda_0^P)$ where 
$\ln \Lambda_0^P$, $\ln \Lambda_0^A$ are the largest Lyapunov exponents of the 
TM, respectively with periodic and antiperiodic boundary conditions across.

The duality between correlation length $\xi$ and
interface tension $\sigma$ is well-established~\cite{watson} for pure 
two-dimensional systems, and carries over to disordered cases. In a 
finite-size scaling (FSS) context~\cite{barber}, this means that 
$\sigma_L$ must scale with $1/L$ at criticality, a fact which
has been used in previous studies of disordered systems~\cite{mm84}, 
including investigations of the NP~\cite{hpp01,dq06,php06}.
From  conformal invariance~\cite{car84} one has, at the critical point:
\begin{equation}
L\,\sigma_L = \pi\eta\ ,
\label{eq:sigeta}
\end{equation}
where, for pure systems, $\eta$ is the same exponent which characterizes
the decay of spin-spin correlations. In the presence of disorder, however, 
the scaling indices of the disorder correlator (i.e., the interfacial tension)
differ from those of its dual, the order correlator (namely, spin-spin
correlations)~\cite{mc02b}. Nevertheless, the constraints of conformal
invariance still hold, thus the amplitude of the domain wall
energy remains a {\em bona fide} universal quantity~\cite{mc02b}. For the
NP, recent estimates on the SQ lattice ($\pm J$ couplings) give 
$\eta=0.691(2)$~\cite{hpp01,mc02a,mc02b}.

We have calculated $\Lambda_0^P$, $\Lambda_0^A$ for strips of SQ, T and HC 
lattices.
Recalling that both 
$L$ in Eq.~(\ref{eq:sigeta}) and the 
correlation length $\xi$ (of which the surface tension is the dual) are actual 
physical distances in lattice parameter units~\cite{pf84,bww90,bn93,dq00}, 
one finds (see Ref.~\onlinecite{dq06}) that, in terms of the number of sites $N$ 
across the strip, the appropriate expressions for the scaled domain-wall energy
are of the form:
$\eta_N =\eta_N(T,z) =\zeta_i\,(N/\pi)(\left(\ln\Lambda_0^P (T,z)
-\ln \Lambda_0^A (T,z)\right)$, with $\zeta_i=1,\, 2/\sqrt{3},\, \sqrt{3}/2$
respectively for $i=$ SQ, T, HC, and where $z=p$ ($\pm J$) or $J_0$ (Gaussian).
At $(T_c,z_c)$ one must have $\lim_{N\to\infty} \eta_N=\eta$, the latter
being a universal quantity. 

Close to the multicritical NP, the scaling directions are respectively
the NL itself, and the temperature axis~\cite{ldh88,ldh89,has08}.  
Therefore (neglecting corrections to scaling), along the NL the single 
relevant variable corresponds to  $z-z_c$.

According to finite-size scaling~\cite{barber}, the curves of scaled domain-wall
energy calculated for different values of $N, T, z$ along the NL should then 
coincide when plotted against $x \equiv N^{1/\nu}\,(z-z_c)$.
  
Bearing in mind that corrections to scaling may be present~\cite{php06,has08},
we allow for their effect from the start. Thus, we write~\cite{has08}:
\begin{equation}
\eta_N=f [ N^{1/\nu}\,(z-z_c)]+ N^{-\omega} 
g [N^{1/\nu}\,(z-z_c) ]\ ,
\label{eq:fss}
\end{equation}
where $\omega >0$ is the exponent associated to the leading irrelevant 
operator. Close enough to the NP the scaling functions 
in Eq.~(\ref{eq:fss}) should be amenable to Taylor expansions. One has:
\begin{equation}
\eta_N=\eta+ \sum_{j=1}^{j_m} a_j\,(z-z_c)^j\,N^{j/\nu}+
N^{-\omega}  \sum_{k=0}^{k_m} b_k\,(z-z_c)^k\,N^{k/\nu}\ .
\label{eq:fss2}
\end{equation}
We adjusted our TM data to Eq.~(\ref{eq:fss2}), by means of multiparametric 
nonlinear least-squares fits. The goodness of fit was measured by
the (weighted) $\chi^2$ per degree of freedom 
($\chi^2_{\rm \ d.o.f.}$). We tested several assumptions on  
$k_m$, $j_m$, $\omega$, via their effect on:  (i) the resulting 
$\chi^2_{\rm \ d.o.f.}$, (ii) the stability of the final estimates
for $z_c$, $\eta$, $1/\nu$, and  (iii) the broad compatibility
of estimates for $\eta$ and $1/\nu$  with existing results for assorted
two-dimensional lattices and coupling distributions (under the reasonable
assumption  of universality, which is however provisional, and must be weighted
against the bulk of available evidence).

We found that:\par\noindent
(1) a parabolic form, $j_m=2$, is adequate for the description of the
broad features of data, similarly to
conclusions from the Monte-Carlo study of Ref.~\onlinecite{has08};\par\noindent
(2) Neglecting corrections to scaling (all $b_k \equiv 0$) 
generally gave a $\chi^2_{\rm \ d.o.f.}$ 
at least one order of magnitude larger than
if such corrections are incorporated;\par\noindent
(3) Fixing $k_m=0$ and allowing $\omega$ to vary gave a final
estimate $\omega \sim 0.1-0.2$, which is too low to qualify as a
{\em bona fide} correction-to-scaling exponent; the same happens if one allows
$k_m \geq 1$ with a variable $\omega$; \par\noindent
(4) For fixed $\omega$, using $k_m=1$ reduces the $\chi^2_{\rm \ d.o.f.}$
by a factor of 2--3 compared with making $k_m=0$,
while no noticeable improvement is forthcoming from allowing $k_m >1$,
again in line with Ref.~\onlinecite{has08};\par\noindent
(5) For fixed $\omega$ between $1$ and $2$, results for 
$\eta$ and $1/\nu$  are in fair accord with point (iii) above;
; also, for this range of $\omega$,
$\chi^2_{\rm \ d.o.f.}$ is minimized, at $\simeq 0.1-0.2$, 
compared to any alternative combination of fixed and variable parameters described 
in this paragraph. The coexistence of these facts indicates that,
within the assumed scenario of describing corrections to scaling via a single
(effective) exponent, the range of $\omega$ just quoted is the one that optimizes   
a universality-consistent picture.

Thus, we kept $j_m=2$, $k_m=1$, allowing $1 \lesssim \omega \lesssim 2$ in what
follows. Results for $\omega=1.5$ are shown in Table~\ref{tpc}. 

\begin{table}
\caption{\label{tpc}
TM estimates of critical quantities $z_c$ ($z=p$, $J_0$), $1/\nu$, 
and $\eta$  for lattices and coupling distributions
[binary ($\pm J$) and Gaussian (G)] specified in column 1. Column 2 gives 
conjectured values of $z_c$; quotations from Refs.~\onlinecite{nn02,mnn03,no06},
and (for T and HC lattices with G distribution) Ref.~\onlinecite{ohzeki08}
refer to original conjecture, while  all others refer to improved 
conjecture. All fits used $\omega=1.5$ (fixed), see 
Eq.~(\protect{\ref{eq:fss2}}) and text.
}
\vskip 0.2cm
\begin{ruledtabular}
\begin{tabular}{@{}lccccc}
Type & conj. & $p_c$, $J_{0c}$ & $1/\nu$ & $\eta$ & 
$\chi^2_{\rm \ d.o.f.}$\\
SQ, $\pm J$ & $0.889972$~\protect{\cite{nn02,mnn03}} & $0.89061(6)$ & $0.64(2)$ & 
$0.689(2)$ & $15/116$ \\
{\ } & $0.89079(6)$\footnote{average over four values from 
improved conjecture, Ref.~\protect{\onlinecite{ohzeki08}}}
 & {\ } &{\ } &{\ } &{\ }  \\
SQ, G & $1.021770$~\protect{\cite{nn02,mnn03}} & $1.0193(3)$ & $0.65(3)$ & 
$0.680(2)$ & $28/133$\\
{\ } & $1.021564$~\protect{\cite{ohzeki08}} & {\ } &{\ } &{\ } &{\ }  \\
T, $\pm J$ &  $0.835806$~\protect{\cite{no06}} & $0.83583(6)$ & $0.65(2)$  
& $0.691(2)$ & $18/119$ \\
{\ } & $0.83597(2)$\footnote{average over two values from 
improved conjecture, Ref.~\protect{\onlinecite{ohzeki08}}}
 & {\ } &{\ } &{\ } &{\ }  \\
HC, $\pm J$ & $0.932704$~\protect{\cite{no06}}  & $0.93297(5)$ & $0.65(1)$ 
& $0.702(2)$ & $15.5/79$ \\
{\ } & $0.93260(1)$$^b$
 & {\ } &{\ } &{\ } &{\ }  \\
T, G &  $0.798174$~\protect{\cite{ohzeki08}} & $0.7971(2)$ & $0.66(2)$ & $0.689(2)$ & 
$17/127$ \\
HC, G & $1.270615$~\protect{\cite{ohzeki08}} & $1.2689(3)$ & $0.64(3)$ & $0.690(2)$ & 
$11/87$ \\
\end{tabular}
\end{ruledtabular}
\end{table}
Since the error
bars quoted in the Table only reflect uncertainties intrinsic to the fitting
procedure, we now illustrate (see Table~\ref{tpc2} below) the 
quantitative effects of relaxing 
some of the assumptions specified above. This is especially important as regards
$z_c$, whose calculated fractional uncertainty is 
one to two orders of magnitude smaller
than those for $1/\nu$, $\eta$. Additional checks on the robustness of such 
narrow error bars are therefore in order.

For instance, considering the T, $\pm J$ case, fixing $\omega =1$,~$2$
gives respectively $p_c= 0.83611(8)$, $0.83565(6)$, with $\chi^2_{\rm \ d.o.f.}$ 
varying by less than $10\%$ against its value for $\omega=1.5$.
Overall, it seems that a realistic error bar should at least include the 
fitted values of $z_c$ obtained for $\omega =1$  and $2$. Table~\ref{tpc2}
shows such estimates, denoted by $z_c^{\rm ave}$, where the associated
uncertainties reflect the spread between
these extreme values (their own intrinsic uncertainties generally being
somewhat smaller, see above and Table~\protect{\ref{tpc}}). A remarkable exception
is the SQ, G case, for which the estimate of $J_{0c}$ is virtually unchanged
as $\omega$ varies in the range described. This instance is also an exception in
that the amplitude of the correction term $b_0$ (column 3 of the Table) 
is much smaller than for all
other cases; consequently, neither $J_{0c}$ nor the  
$\chi^2$  (resp. columns 4 and 5) change appreciably when 
corrections to scaling are ignored. The latter is not true for any of the other cases
studied.
\begin{table}
\caption{\label{tpc2}
For lattices and coupling distributions
[binary ($\pm J$) and Gaussian (G)] specified in column 1, column 2 gives
critical quantities $z_c^{\rm ave}$ ($z=p$, $J_0$), averaged over values
from fits for $\omega=1$ and $2$ (see text).  
Coefficients $b_0$, $b_1$ (see Eq.~(\protect{\ref{eq:fss2}}))
fitted for $\omega=1.5$; the index $(0)$ for the last two columns
denotes quantities obtained from fits where corrections to scaling were neglected.
}
\vskip 0.2cm
\begin{ruledtabular}
\begin{tabular}{@{}lccccc}
Type & $z_c^{\rm ave}$ & $b_0$  & $b_1$ & $z_c^{(0)}$ & 
$\chi^{2\,(0)}_{\rm \ d.o.f.}$\\
SQ, $\pm J$ &  $0.89065(20)$ & $-0.126(3)$ & $-3.7(4)$ & $0.8898(1)$
& $416/118$ \\
SQ, G &  $1.0193(4)$ & $0.009(3)$ & $-0.35(18)$ & 
$1.0195(1)$ & $30/135$\\
T, $\pm J$ &  $0.83588(23)$ & $-0.145(3)$ & $-1.4(4)$  & $0.8348(1)$
 & $520/121$ \\
HC, $\pm J$ & $0.93300(15)$  & $-0.142(3)$ & $-2.5(4)$ & $0.9322(1)$
 & $499/81$ \\
T, G &  $0.7972(6)$ & $-0.106(3)$ & $-0.31(18)$ & $0.7948(3)$  & $163/129$ \\
HC, G & $1.2691(10)$ & $-0.152(4)$ & $-0.64(15)$ & $1.2635(9)$ & $325/89$ \\
\end{tabular}
\end{ruledtabular}
\end{table}

Our assessment of the estimates quoted in Table~\ref{tpc2} 
for the location of the NP is as
follows. 

For SQ, $\pm J$ our results are in agreement with the improved
conjecture~\cite{ohzeki08}, and with numerical data from 
Refs.~\onlinecite{hpp01,mc02a,php06,has08}. For T, $\pm J$ our
range of estimates is roughly consistent with the conjecture, both in its
original~\cite{no06} and improved~\cite{ohzeki08} versions. It is also 
at the upper limit of the early estimate $p_c=0.8355(5)$~\cite{dq06}.

For all remaining cases, our numerical data indicate that the conjecture
fails to hold, albeit by rather small amounts, $0.2\%$ at most.
For the $\pm J$ distribution, our results for both
SQ and HC indicate that the conjectured position of the NP lies in the
paramagnetic phase (for SQ, this is true only for the original conjecture).
On the other hand, for the Gaussian distribution and all three lattices, 
according to our estimates the conjecture places the NP slightly inside the 
ordered phase. 

For  HC, $\pm J$ the result in Table~\ref{tpc2} is again at the upper
end of the range given in Ref.~\onlinecite{dq06}, $p_c=0.9325(5)$.
Note also that our estimate for SQ, G lies farther from the conjecture
than the numerical value given in Ref.~\onlinecite{php06}, namely
$J_{0c}=1.02098(4)$ (thus, this latter also places the conjectured
location of the NP inside the ordered phase).  
The above estimates of $p_c$ and $J_{0c}$ for T and HC lattices, when plugged into 
Eq.~(\ref{eq:earlyconj}), using Eq.~(\ref{eq:h(p)}) and its counterpart for 
Gaussian distributions~\cite{ohzeki08}, result in:
\begin{eqnarray}
H(p_{1c}) + H(p_{2c}) =0.9986(12)\qquad\ \  (\pm J)\ ;
\nonumber \\
H(J_{0c1}) + H(J_{0c2}) =1.0014(10)\quad ({\rm Gaussian})\ ,
\label{eq:hest}
\end{eqnarray} 
both narrowly missing the conjecture of Eq.~(\ref{eq:earlyconj}). 

As regards the correlation-length exponent and the critical amplitude $\eta$, we 
found that, for each lattice and coupling distribution, the error bars quoted
in Table~\ref{tpc} are wide enough to accommodate the variations in central estimates,
both when one sweeps $\omega$ between $1$ and $2$ as above, and when $z_c$ is varied
between the limits established in Table~\ref{tpc2}. No evidence emerges from the data
which justifies challenging our earlier assumption of universality. 
From  unweighted averages over the respective columns of Table~\ref{tpc}, we quote
$\nu=1.53(4)$, $\eta=0.690(6)$. These are to be compared to the recent results
$\nu=1.50(3)$~(SQ, Ref.~\onlinecite{mc02a}), $1.48(3)$~(SQ, Ref.~\onlinecite{php06}),
$1.49(2)$~(T and HC, Ref.~\onlinecite{dq06}), $1.527(35)$~(SQ, Ref.~\onlinecite{has08}), 
all for $\pm J$ distributions; see also $\nu=1.50(3)$~(SQ, Ref.\onlinecite{php06}), 
Gaussian. For the critical amplitude, we recall (all for  $\pm J$):
$\eta=0.691(2)$~\cite{hpp01,mc02a,mc02b} (SQ); $0.674(11)$ (T), $0.678(15)$ (HC), both
from Ref.~\onlinecite{dq06} .

\begin{figure}
{\centering \resizebox*{3.4in}{!}
{\includegraphics*{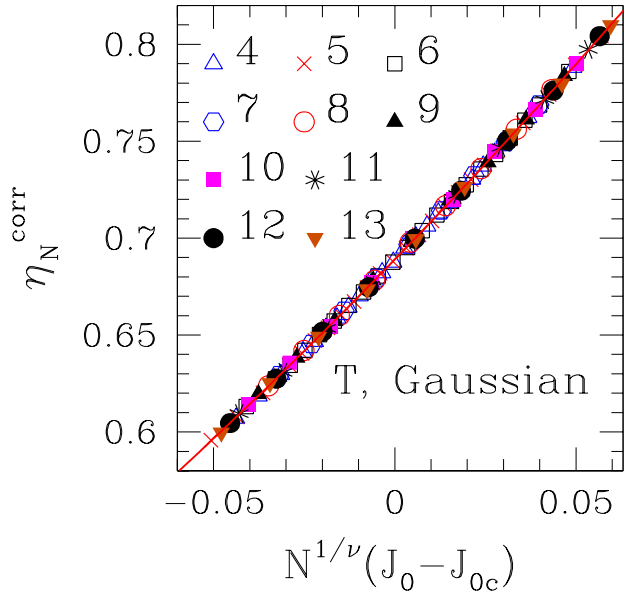}} 
\par}
\caption{(Color online) Triangular lattice, Gaussian coupling distribution: 
scaling plot of domain-wall free energies,
subtracting corrections to scaling: 
$\eta_N^{\rm corr} =\eta_N -N^{-\omega} g(N^{1/\nu}(J_0-J_{0c}))$ [$\,$see 
Eq.~(\protect{\ref{eq:fss}})$\,$], against the scaling variable 
$N^{1/\nu}(J_0-J_{0c})$.
Error bars are smaller than symbol sizes. The full line is a quadratic fit to
corrected data.
}
\label{fig:dwtsc}
\end{figure}

\begin{figure}
{\centering \resizebox*{3.4in}{!}
{\includegraphics*{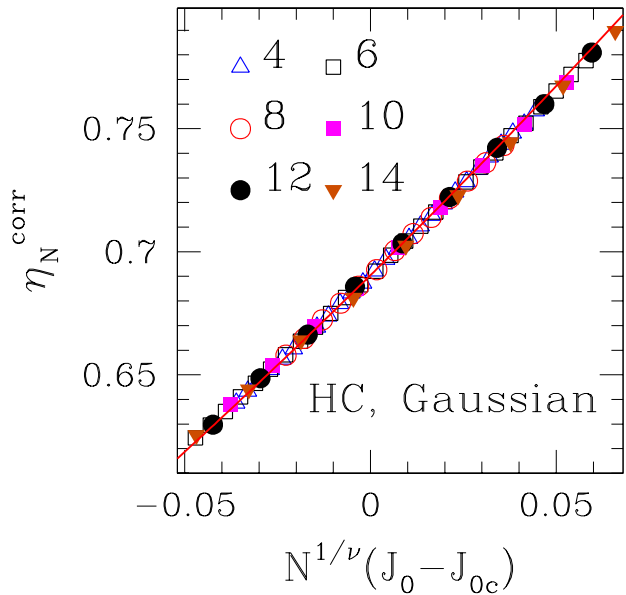}} 
\par}
\caption{(Color online) Honeycomb lattice, Gaussian coupling distribution: 
scaling plot of domain-wall free energies,
subtracting corrections to scaling: 
$\eta_N^{\rm corr} =\eta_N -N^{-\omega} g(N^{1/\nu}(J_0-J_{0c}))$ [$\,$see 
Eq.~(\protect{\ref{eq:fss}})$\,$], against the scaling variable 
$N^{1/\nu}(J_0-J_{0c})$.
Error bars are smaller than symbol sizes. The full line is a quadratic fit to
corrected data.
}
\label{fig:dwhcsc}
\end{figure}

The overall quality of our scaling plots is illustrated in 
Figures~\ref{fig:dwtsc}
and~\ref{fig:dwhcsc}. We chose to display data for T and HC, Gaussian 
distributions, because  for these there are 
fewer data available in the literature. As the last
column of Table~\ref{tpc} shows, the $\chi^2_{\rm d.o.f.}$ remains
very much in the same neighborhood for all cases studied.


\section{Central charge}
\label{sec:2a}
We used the free-energy data generated in
Section~\ref{sec:2} also to estimate the conformal anomaly, or central charge 
$c$, at the NP. This is evaluated via the finite-size scaling of the 
free energy on a strip with periodic boundary conditions across~\cite{bcn86},
\begin{equation}
 f(T_{c},N)=f(T_{c},\infty)+\frac{\pi c}{6N^{2}}+ \frac{d}{N^{4}}+{\cal 
O}\left(\frac{1}{N^6}\right) \label{eq:cc}
\end{equation}
where $f(T_{c},\infty)=\lim_{L\rightarrow \infty}\,f(T_{c},L)\,$
is a regular term which corresponds to the bulk system free energy.
For disordered systems, Eq.~(\ref{eq:cc}) is expected to hold when
the configurationally-averaged free energy is considered, with
$c$ taking the meaning of an \emph{effective} conformal
anomaly~\cite{lc87,jc98,php06}~.

By writing only even powers of $N^{-1}$ in Eq.~(\ref{eq:cc}), 
it is assumed that only analytic
corrections come up~\cite{car86}. While this is true, e.g., for
pure Ising systems, a counterexample is the three-state Potts ferromagnet
for which  free-energy corrections in $N^{-2\omega_0}$, $N^{-3\omega_0}\ \dots$, 
$\omega_0=4/5$, are present~\cite{nienhuis82,dq00}. Although not much
is known about the operator structure at the NP, existing central charge 
estimates in this case have been derived via  Eq.~(\ref{eq:cc})
so far with fairly consistent results, namely c=0.464(4)~\cite{hpp01,php06},
0.46(1)~\cite{ldq06}. We shall return to this point at the end of
this Section.

We have evaluated free energies at the predicted locations of the NP given in
Table~\ref{tpc2}, both at the central estimates and at either end of the respective 
error bars. We found that such values can be calculated with sufficient accuracy,
via interpolation of those already computed at the sets of  
equally-spaced points used originally in Section~\ref{sec:2}. 
Results for the central charge are displayed in Table~\ref{tcc}, where error 
bars for all quantities mostly reflect uncertainties intrinsic to the
fitting procedure itself, as our estimates are    
rather stable along the predicted intervals of location of the NP.
Indeed, it is expected~\cite{php06} that at 
criticality the  calculated conformal anomaly passes through a maximum as a 
function of position along the NL. 

\begin{table}
\caption{\label{tcc}
Conformal anomaly $c$ and non-universal higher-order coefficent $d$, from
fits of critical free-energy data to Eq.~(\protect{\ref{eq:cc}}).
Last column gives fitted values of $c$ under the asssumption that $d \equiv 0$
in Eq.~(\protect{\ref{eq:cc}}).
}
\vskip 0.2cm
\begin{ruledtabular}
\begin{tabular}{@{}lccc}
Type &  $c$ & $d$ & $c\,[d \equiv 0]$\\
SQ, $\pm J$ & $0.463(3)$ & $0.13(1)$ &   $0.478(2)$\\
SQ, G &  $0.461(4)$  & $0.14(3)$& $0.476(2)$\\
T, $\pm J$ & $0.459(3)$  & $0.01(1)$ & $0.461(1)$ \\
HC, $\pm J$ & $0.457(5)$   & $0.02(2)$ &  $0.462(2)$ \\
T, G &  $0.454(4)$ & $0.06(3)$ & $0.461(1)$\\
HC, G &  $0.468(15)$  & $-0.05(6)$& $0.459(5)$ \\
\end{tabular}
\end{ruledtabular}
\end{table}

In line with earlier findings~\cite{ldq06}, one sees that for SQ and both 
coupling 
distributions, ignoring the fourth-order term in Eq.~(\ref{eq:cc}) 
shifts the final estimate of $c$ by some $4-5$ error bars, away from the 
expected universal value $\sim 0.46$.
On the other hand, for T and HC the fitted  $d$ is much closer to zero
than for SQ; furthermore, for these latter lattices, results obtained fixing $d=0$
appear generallly more consistent with universality, and with less spread,
than those found with $d$ kept as a free parameter. Overall, 
we interpret the above results as indicating that: (i) there is no 
evidence for universality breakdown as regards the conformal anomaly; taking this 
as true, (ii) there appears to be no unusual (non-analytic) free-energy scaling 
correction  $N^{-\omega_0}$ with $2 < \omega_0 < 4$; and (iii) it is possible
that the fourth-order term  is $d \equiv 0$ for T and HC, similarly to the case
of pure Ising systems~\cite{dq00}.

\section{Uniform susceptibilities}
\label{sec:susc}
We calculated uniform zero-field susceptibilities along the NL, for SQ, T and HC
lattices and only for Gaussian  distributions, as done in previous 
investigations for $\pm J$~\cite{dq06,sbl3}. 
For the finite differences used in numerical differentiation, we used a field step
$\delta h=10^{-4}$ in units of $T$. We swept the same respective intervals of 
$J_0$ quoted in Table~\ref{tpar}.

Finite-size scaling arguments~\cite{barber} suggest a form
\begin{equation}
\chi_N = N^{\gamma/\nu}\,f [\,N^{1/\nu}(J_0-J_{0c})\,]\ ,
\label{eq:chisc}
\end{equation}
where $\chi_N$ is the  finite-size susceptibility, and $\gamma$ is the susceptibility
exponent. In order to reduce the number of fitting parameters, we kept $1/\nu$ 
and $J_{0c}$
fixed at their central estimates obtained in Sec.~\ref{sec:2}, and allowed 
$\gamma/\nu$ to vary. Again, we took corrections to scaling into account.  
Following Ref.~\onlinecite{has08}, we write:
\begin{eqnarray}
\ln \chi =\frac{\gamma}{\nu} \ln N +
\sum_{j=1}^{j_m} a_j\,(J_0-J_{0c})^j\,N^{j/\nu}+
\nonumber \\
+ N^{-\omega}  \sum_{k=0}^{k_m} b_k\,(J_0-J_{0c})^k\,N^{k/\nu}\ .
\label{eq:fss3}
\end{eqnarray}
Similarly to Section~\ref{sec:2} above, we found that choosing $j_m=2$, $k_m=1$
enables one to obtain good fits to numerical data, with $\chi^2_{\rm d.o.f.} \simeq 
01.-0.2$. The consequences of keeping $\omega$ as a free parameter or, on the other
hand, fixing its value during the fitting procedure, can be seen in 
Table~\ref{tchi}.
\begin{table}
\caption{\label{tchi}
For the zero-field susceptibility and lattices as specified in column 1 (all 
with Gaussian coupling distributions),
columns 2, 3 , 4 give the leading correction-to scaling exponent as fitted,
and the corresponding $\gamma/\nu$ and $\chi^{2}_{\rm \ d.o.f.}$; columns 5, 6
give the two latter quantities, now taken by keeping $\omega$
fixed during the fitting procedure, and averaging over resulting 
values for $\omega=1$ and $2$ (see text).
}
\vskip 0.2cm
\begin{ruledtabular}
\begin{tabular}{@{}lccccc}
Type & $\omega^{\rm fit}$ & $\gamma/\nu$ & $\chi^{2}_{\rm \ d.o.f.}$ & 
$(\gamma/\nu)^{\rm ave}$ & $\chi^{2\, {\rm ave}}_{\rm \ d.o.f.}$\\
SQ, G &  $1.3(3)$ & $1.79(2)$ & $0.127$ & 
$1.793(6)$ & $0.129$\\
T, G & $0.7(3)$  &  $1.81(1)$ & $0.118$ & $1.814(6)$  & $0.128$ \\
HC, G & $0.4(6)$ & $1.79(4)$ & $0.17$ & $1.804(7)$ & $0.18$ \\
\end{tabular}
\end{ruledtabular}
\end{table}
While the fitted value of $\omega$ for SQ looks acceptable, the same cannot be said of 
that for HC, as the associated error bar allows even slightly negative values 
(the result for T being half-way between the other two). Also, 
by  keeping $\omega$ as a free parameter, one gets an error bar for 
$\gamma/\nu$ that is at least twice that obtained if $\omega$ is kept fixed between $1$ 
and $2$, without any noticeable improvement in the $\chi^{2}_{\rm \ d.o.f.}$.
On the other hand, using fixed $\omega$ above this latter range results in a slow but 
steady loss of quality: for example, for the T lattice, $\omega=4$ gives 
$\chi^{2}_{\rm \ d.o.f.} \simeq 0.2$. 
Thus, although the idea of allowing $\omega$ to vary freely seems, in principle,
the correct thing to do, the results in this particular case do not appear to
be obviously more reliable than those averaged for fixed $\omega$ between $1$ and $2$.
We then decided to use these latter as our main reference.
Taking an unweighted average over the three  estimates for $(\gamma/\nu)^{\rm ave}$ 
gives the final value $\gamma/\nu=1.804(16)$. This is to be compared to the following
(all for $\pm J$ distributions): $1.80(2)$~\cite{sbl3} [SQ]; $1.795(20)$ [T] and 
$1.80(4)$ [HC], both from Ref.~\onlinecite{dq06}; $1.80$--$1.82$\cite{php06} [SQ].

\section{Correlation functions}
\label{sec:cf}
Our study of correlation functions is based on previous work for SQ~\cite{dqrbs03}, 
T and HC lattices~\cite{dq06} ($\pm J$ only). In this Section, we specialize to
the Gaussian distribution, for T and HC lattices only.
On the NL, the moments of the 
PDF for the correlation function between Ising
spins $\sigma_i$, $\sigma_j$ are equal two by two~\cite{nish81,nish01,nish86,nish02}:
\begin{equation}
[\,\langle \sigma_i \sigma_j\rangle^{2\ell -1}] =
[\,\langle \sigma_i \sigma_j \rangle^{2\ell} ]\ ,
\label{eq:momsc}
\end{equation}
where angled brackets indicate thermal average,  square brackets
stand for configurational averages over disorder, and $\ell = 1, 2, \dots$.

At the NP, conformal invariance~\cite{cardy87} is expected to hold, 
provided suitable averages over disorder are 
considered~\cite{ludwig90,dq95,dq97,hpp01,mc02a,mc02b,dqrbs03,dq06}.
On a strip of width $L$ of a square lattice, with periodic boundary conditions
across, the  disorder-averaged $k$-th moment of the correlation 
function PDF between spins located respectively at the origin and at $(x,y)$ 
behaves at criticality as:
\begin{equation}
[\,\langle \sigma_i \sigma_j\rangle^{k}] \sim z^{-\eta_k}\ ,\ 
z \equiv [\,\sinh^2 (\pi x/L)+ \sin^2 (\pi y/L)\,]^{1/2}\  .
\label{eq:conf-inv}
\end{equation}
For the T and HC lattices, the same is true, provided that the actual, i.e.,
geometric site coordinates along the strip are used in Eq.~(\ref{eq:conf-inv}).
Details are given in Ref.~\onlinecite{dq06}.
Note that Eq.~(\ref{eq:momsc}) implies $\eta_{\,2\ell-1}=\eta_{\,2\ell}$.

As in earlier work~\cite{dqrbs03,dq06}, we concentrate on short-distance correlations, 
i.e., where the  argument $z$ is strongly influenced by $y$. Such a setup is especially 
convenient in  order to probe the angular dependence predicted in 
Eq.~(\ref{eq:conf-inv}), which
constitutes a rather stringent test of conformal invariance properties.

Following Refs.~\onlinecite{hpp01,dqrbs03,dq06}, we extract the  
decay-of-correlations exponents $\eta_{\,k}$, via least-squares fits of our
data to the form $m_{\,k} \sim z^{-\eta_{\,k}}$. We also consider the 
exponent $\eta_0$ which characterizes the zeroth-order 
moment of the correlation-function distribution~\cite{ldq06}, 
i.e. it gives the typical, or most probable, value of this
quantity (see, e.g., Ref.~\onlinecite{dq97} and references therein). 
One has, in the bulk,
\begin{equation}
G_0 (R) \equiv \exp\left[ \ln \langle  \sigma_{0}\sigma_{R} \rangle \right]_{\rm av}
\sim R^{-\eta_0}\ .
\label{eq:eta0}
\end{equation}
Calculations on strips of the $\pm J$ SQ lattice, at the early conjectured location 
of the NP~\cite{nn02}, gave the estimate $\eta_0=0.194(1)$~\cite{ldq06}~.

As seen in earlier work~\cite{dqrbs03}, for strip widths $N=10$ or 
thereabouts,  finite-width effects are already mostly subsumed in the explicit 
$L$ (i.e., $N$) dependence of Eq.~(\ref{eq:conf-inv}). 
However, some detectable (albeit tiny)  variations in the calculated values of
averaged moments of the correlation function PDF may still 
persist upon varying $N$.
These are of course minimized at the critical point where the bulk correlation length 
diverges. We have calculated correlation functions for $N \leq 12$, 
for values of $J_0$ within the error bars given for the location of the NP in 
Table~\ref{tpc2}~. We have seen that along
these intervals of $J_0$, the trend followed by such averaged moments against 
$N$-- variation is as follows: for T, it cannot be distinguished from stability
within error bars, 
while for HC it is slightly downward (of the order of one error bar from $N=10$ 
to $N=12$). 
For fixed $N$ and $J_0$, one error bar associated to intrinsic fluctuations 
is $\lesssim 1\%$.

\begin{table}
\caption{\label{teta}
Estimates of exponents $\eta_{\,k}$, from least-squares fits of
averaged moments of correlation-function distributions to  the form
$m_{k} \sim z^{-\eta_{\,k}}$, for $z$ defined in 
Eq.~(\protect{\ref{eq:conf-inv}}). For columns 2 and 3 (this work), central  
estimates and 
error bars reflect averages between results for $N=10$ and $12$, as well as
variations from scanning $J_0$ along the error bars for locations of NP given in 
Table~\protect{\ref{tpc2}}. Columns 4, 5, 6 quote existing data for comparison.
For SQ, all results are for $\pm J$ coupling distribution, unless otherwise
noted.
}
\vskip 0.2cm
\begin{ruledtabular}
\begin{tabular}{@{}lccccr}
$k$ & T (G)      & HC (G)     & T ($\pm J$)~\protect{\cite{dq06}} &  HC ($\pm 
J$)~\protect{\cite{dq06}} &   SQ{\qquad } \\
$0$ & $0.185(3)$ & $0.184(3)$ &    --       &       --      &  
$0.194(1)$~\protect{\cite{ldq06}}   \\
1   & $0.178(2)$ & $0.178(2)$  &  $0.181(1)$ &  $0.181(1)$   & 
$0.1854(17)$~\protect{\cite{dqrbs03}}   \\ 
{\ } & {\ } & {\ } & {\ } & {\ } &  $0.1854(19)$~\protect{\cite{hpp01}} \\
{\ } & {\ } & {\ } & {\ } & {\ } &  $0.183(3)$~\protect{\cite{mc02a}} \\
{\ } & {\ } & {\ } & {\ } & {\ } &  $0.1848(3)$~\protect{\cite{php06}} \\
{\ } & {\ } & {\ } & {\ } & {\ } &  $0.1818(2)$ [G]~\protect{\cite{php06}} \\
{\ } & {\ } & {\ } & {\ } & {\ } &  $0.180(5)$~\protect{\cite{has08}} \\
3 &  $0.250(2)$  & $0.252(2)$ & $0.251(1)$  & $0.252(1)$   & 
$0.2556(20)$~\protect{\cite{dqrbs03}}   \\ 
{\ } & {\ } & {\ } & {\ } & {\ } &  $0.2561(26)$~\protect{\cite{hpp01}} \\
{\ } & {\ } & {\ } & {\ } & {\ } &  $0.253(3)$~\protect{\cite{mc02a}} \\
{\ } & {\ } & {\ } & {\ } & {\ } &  $0.2552(9)$~\protect{\cite{php06}} \\
{\ } & {\ } & {\ } & {\ } & {\ } &  $0.2559(2)$ [G]~{\cite{php06}} \\
5 &  $0.296(2)$    & $0.300(5)$  &  $0.297(2)$ &  $0.296(2)$ &  
$0.300(2)$~\protect{\cite{dqrbs03}}   \\
{\ } & {\ } & {\ } & {\ } & {\ } &  $0.3015(30)$~\protect{\cite{hpp01}} \\
{\ } & {\ } & {\ } & {\ } & {\ } &  $0.3004(13)$~\protect{\cite{php06}} \\
{\ } & {\ } & {\ } & {\ } & {\ } &  $0.3041(2)$ [G]~\protect{\cite{php06}} \\
7 &  $0.331(4)$   & $0.336(6)$   &  $0.330(2)$ & $0.329(3)$ & 
$0.334(3)$~\protect{\cite{dqrbs03}}   \\
{\ } & {\ } & {\ } & {\ } & {\ } &  $0.3354(34)$~\protect{\cite{hpp01}} \\
{\ } & {\ } & {\ } & {\ } & {\ } &  $0.3341(16)$~\protect{\cite{php06}} \\
{\ } & {\ } & {\ } & {\ } & {\ } &  $0.3402(2)$ [G]~\protect{\cite{php06}} \\
\end{tabular}
\end{ruledtabular}
\end{table}

Table~\ref{teta}  gives numerical results of the fits for $k=0$, and odd 
$k>1$ (we have also calculated even moments for $k \geq 2$ and checked that
Eq.~(\ref {eq:momsc}) holds). 
One sees that T and HC estimates are quite consistent 
with each other for all $k$. On the other hand, for $k=0$, $1$, and $3$ 
they fall slightly below their 
existing counterparts, given in columns 4, 5, and 6 of the Table.  
For $k=5$ and $7$, as a consequence of generally wider error bars, all estimates 
are broadly compatible with one another.

Physically, obtaining (via least-squares fits) a smaller [larger] than expected 
value for the decay-of-correlations exponent
would indicate that it is being evaluated inside the ordered 
[paramagnetic] phase, instead of right at the critical point. 

Applying these ideas to the $k=0$ case, we recall that the result of 
Ref.~\onlinecite{ldq06} for SQ, $\pm J$ was calculated at the originally 
conjectured position of the NP~\cite{nn02}. By now, it seems well established 
that this point is in the disordered phase (see Table~\ref{tpc}). Therefore,
the value of Ref.~\onlinecite{ldq06} should be taken as an upper bound,
which is obeyed by our present estimates.

For $k=1$ and $3$, one might use the same argument as above to argue that
the result of Ref.~\onlinecite{dqrbs03} is too large, as it was calculated
at the same point as that of Ref.~\onlinecite{ldq06}. On the other hand, this
cannot be said of the additional estimates quoted in the Table, all of which
are also larger than ours (though in some cases the respective error bars
overlap, or at least touch each other). Using the 
reasoning described above, one would infer
that for T and HC with Gaussian distribution, the ranges of locations 
for the NP given in Table~\ref{tpc2} are in fact both inside the ordered phase.
Since these latter, in their turn, put the conjectured NP position also
inside the ordered phase, the final conclusion would be that the actual
location of the NP differs from the conjecture by an  amount larger than
predicted by domain-wall (DW) scaling: $J_{0c}^{\,\rm real} < 
J_{0c}^{\,\rm DW} < J_{0c}^{\,\rm conj.}$. The slight downward trend against increasing 
$N$, reported above for HC, would be consistent with this scenario. However,
we have not seen a similar trend for T. 

One should note also that all the discrepancies remarked upon are rather 
small: the single worst case, as regards central estimates, is that of 
the present result  $\eta_1=0.178$ against $\eta_1=0.1854$~\cite{hpp01,dqrbs03}, 
amounting to $4\%$, or $\simeq 3.5$ times the respective error bar.
Given that the quoted values (especially those for the associated uncertainties)
are likely to depend on details of the respective fitting procedures, the
resulting picture looks mixed. 
 
In conclusion, existing evidence does not
seem strong enough to state that our estimates from Sec.~\ref{sec:2} for the 
location of the NP on T and HC lattices (Gaussian distribution) are definitely
inside the ordered phase.

\section{Discussion and Conclusions}
\label{conc}
We have used domain-wall scaling techniques in Sec.~\ref{sec:2} to determine 
the location of the Nishimori point of Ising spin glasses.
In the analysis of our data we allowed for the existence of corrections to 
scaling, see Eqs.~(\ref{eq:fss}) and ~(\ref{eq:fss2}). 

Results for the SQ lattice, $\pm J$ distribution, show that such corrections
play a crucial role in the finite-size scaling of domain-wall energies.
Indeed, when they are taken into account, the estimated position of the NP
is $p_c=0.89065(20)$, in excelent agreement with recent and very accurate
numerical work~\cite{php06,has08}. One can see from the two last columns of
Table~\ref{tpc2} that, if corrections to scaling are ignored, the value
of $p_c$ which minimizes the $\chi^2_{\rm d.o.f.}$ (though at a level
$\sim 30$  times that obtained when corrections are incorporated) is
instead $p_c=0.8898(1)$, very close to the original conjecture and
incompatible with the above-mentioned body of numerical evidence.
In retrospect, one sees that the domain-wall scaling result of 
Ref.~\onlinecite{dq06} for this case, namely  $p_c=0.8900(5)$, essentially 
suffers from  
the effect of ignoring corrections to scaling (though even so it still
picks out the correct exponent, $1/\nu =1.45(8)$~\cite{dq06}).

Going over to SQ, Gaussian, domain-wall scaling for strips of widths
$4 \leq N \leq$ 14 sites gives $J_{0c}=1.0193(4)$, lower than
both the conjecture (original and improved) and the result of
Ref.~\onlinecite{php06}, $J_{0c}=1.02098(4))$. In that 
Reference, the mapping of the spin problem to a network model (described in
Ref.~\onlinecite{mc02a}) enabled the authors to reach significantly larger
lattice sizes than here. The result just quoted
was obtained by extrapolation of $11 \leq N \leq 24$ crossing-point data, 
without explicit account of corrections to scaling (which, as those authors 
show, do produce a trend reversal around $N = 8$, and are expected to
have negligible effect for the large widths used in the extrapolation).
It may be that our own data fail to incorporate
an underlying trend which only comes about for larger systems. 
Nevertheless, the stability of our results for this particular case is 
remarkable, as pointed out in the initial discussion of Table~\ref{tpc2}.

Our results for T and HC, $\pm J$ distribution, are marginally compatible with, 
but more accurate than, the earlier ones of Ref.~\onlinecite{dq06}; 
though for T they are also
broadly compatible with the conjecture in both original and improved versions,
for HC  our estimate in Table~\ref{tpc2} lies at least two error bars away
from the conjecture.

For  T and HC, Gaussian distribution, in both cases the discrepancy between our
results and the conjecture is again of the order of two error bars.

Consequently, as shown in Eq.~(\ref{eq:hest}), we predict the 
duality-based conjecture of 
Eq.~(\ref{eq:earlyconj}) to be narrowly missed, for both $\pm J$ and Gaussian
cases, though on opposite sides of the hypothesized equality. 

As regards the exponent $\nu$ and critical amplitude $\eta$ (see 
Eq.~(\ref{eq:sigeta})) which are also estimated via domain-wall scaling, 
we have found no evidence of nonuniversal, lattice-- or disorder 
distribution-- dependent, behavior. Therefore, from unweighted averages
over all six cases studied, we quote $\nu=1.53(4)$, $\eta=0.690(6)$.
Both are in very good agreement with existing numerical results
(see the end of Sec.~\ref{sec:2} for detailed comparisons). 

The conformal anomaly values calculated in Sec.~\ref{sec:2a} are 
in good agreement among themselves and with previous 
estimates~\cite{hpp01,php06,ldq06}. Our fits for the non-universal
coefficient $d$ of the fourth-order correction to the critical free energy
suggest that $d \equiv 0$ for T and HC lattices (while definitely $d \neq 0$
for SQ). This would be similar to the lattice-dependent structure of corrections
for pure Ising systems~\cite{dq00}. An unweighted average of values from
Table~\ref{tcc} (using results of fits with $d \neq 0$ for SQ, and with
$d \equiv 0$ for T and HC) gives $c=0.461(5)$.

In Sec.~\ref{sec:susc} we evaluated uniform zero--field susceptibilities,
by direct numerical differentiation of the free energy against external 
field. Only Gaussian distributions were considered, for SQ, T, and HC.
Though our results show some lattice-dependent spread, the error bars
for $(\gamma/\nu)^{\rm ave}$ still overlap in pairs. It is known
that susceptibility calculations are prone to larger fluctuations than, e.g.,
domain-wall energy ones~\cite{php06}. In the Monte Carlo 
simulations of Ref.~\onlinecite{has08}, this effect was reduced by
considering the quantity $\chi/\xi^2$ (where $\xi$ is the finite-size
correlation length), which behaves more smoothly than $\chi$ on its own.
Our final estimate (averaged over results for all three lattices),
$\gamma/\nu=1.804(16)$, compares favorably (albeit somewhat close to 
marginally) with the corresponding 
one from Ref.~\onlinecite{has08}, $\gamma/\nu=1.820(5)$.

Finally, in Sec.~\ref{sec:cf} we applied conformal-invariance concepts
to the statistics of spin-spin correlation functions, extracting the associated 
multifractal scaling  exponents~\cite{hpp01,mc02a,mc02b,php06}. 
We only examined T and HC lattices, for Gaussian coupling distributions. 
The overall  picture summarized in Table~\ref{teta} 
points towards  universality of the exponents $\{ \eta_k \}$, though some small 
discrepancies remain. The case $k=1$ is especially relevant, on account
of its connection with the uniform susceptibility
via the scaling relation $\gamma/\nu=2-\eta_1$. While our result 
$\eta_1=0.178(2)$ is somewhat lower than existing data from direct 
calculations of correlation functions, it gives
$\gamma/\nu=1.822(2)$ when inserted in the scaling relation. 
This agrees very well with the above-quoted estimate~\cite{has08}, 
$\gamma/\nu=1.820(5)$.

In summary, we have produced estimates of the location of the NP on 
SQ, T and HC lattices, and for $\pm J$ and Gaussian coupling distributions.
Though these are consistent with existing conjectures for  SQ and T (both
$\pm J$), they appear to exclude the respective conjectured values for
the remaining cases. However, the discrepancies are very small, amounting
to $0.2\%$ in the worst case (SQ, Gaussian). Furthermore, we have assessed
several critical quantities (amplitudes and exponents), and found an
overall picture consistent with universality as regards lattice
structure and disorder distribution.

\begin{acknowledgments}
The author thanks Hidetoshi Nishimori, Marco Picco, Andrea Pelissetto,
Francesco Parisen Toldin, and Masayuki Ohzeki for many enlightening 
discussions. 
This research  was partially supported by  the Brazilian agencies CNPq 
(Grant No. 30.6302/2006-3), FAPERJ (Grants Nos. E26--100.604/2007 and 
E26--110.300/2007), CAPES, and Instituto do Mil\^enio de Nanotecnologia--CNPq.
\end{acknowledgments}

\bibliography{biblio}  
\end{document}